# FOUR-STATE NANOMAGNETIC LOGIC USING MULTIFERROICS

Noel D'Souza[1,a], Jayasimha Atulasimha[a] and Supriyo Bandyopadhyay[b]

(a) Department of Mechanical and Nuclear Engineering,

(b) Department of Electrical and Computer Engineering,

Virginia Commonwealth University, Richmond, VA 23284, USA.

**Abstract**

The authors show how to implement a 4-state universal logic gate (NOR) using three strain-coupled magnetostrictive-piezoelectric multiferroic nanomagnets (e.g. Ni/PZT) with biaxial magnetocrystalline anisotropy. Two of the nanomagnets encode the 2-state input bits in their magnetization orientations and the third nanomagnet produces the output bit via dipole interaction with the input nanomagnets. A voltage pulse alternating between -0.2 V and +0.2 V is applied to the PZT layer of the third nanomagnet and generates alternating tensile and compressive stress in its Ni layer to produce the output bit, while dissipating ~ 57,000 $kT$ (0.24 fJ) of energy per gate operation.

---

[1]Corresponding author. E-mail dsouzanm@vcu.edu

Nanomagnetic logic (NML) is an emerging paradigm for low-power computing[1-4] where logic bits 0 and 1 are encoded in two stable magnetization directions parallel to the easy axis of a single domain nanomagnet with uniaxial shape anisotropy. Bits are switched by flipping the magnetization of a nanomagnet with an external agent that acts like a clock.

The ultimate energy-efficiency of NML is determined by the clocking methodology, i.e. how bits are flipped. If a magnetic field generated with a current[5], or spin transfer torque[6], is used to flip a nanomagnet's magnetization, then the energy consumed can be exorbitant, which defeats the very purpose of NML. However, stress-induced rotation of magnetization in *multiferroic* nanomagnets seems to be able to switch magnets with much higher energy efficiency[7-9]. This is only now beginning to attract attention[7-10].

The magnetization vector of a multiferroic nanomagnet, consisting of a piezoelectric and a magnetostrictive layer, can be flipped with an electrostatic potential applied to the piezoelectric layer. The voltage generates a strain that is elastically transferred to the magnetostrictive layer causing its magnetization to flip. We can embellish the functionality of the multiferroic element by introducing biaxial *magnetocrystalline* anisotropy in the magnetostrictive layer, giving it four possible stable magnetization directions ('up', 'right' ,'down', 'left') that are chosen to encode four possible 2-bit combinations (00, 01, 11, 10), illustrated in Fig. 1(a). The choice is motivated by allowing only 1 bit to change for every 90⁰ rotation of magnetization. For single crystal Ni, with magnetocrystalline anisotropy constant $K_1 < 0$, the "easy" directions that encode these states in the (001) plane are the [110], [1$\bar{1}$0], [$\bar{1}$$\bar{1}$0] and [$\bar{1}$10] directions, as shown in the energy curves in Fig. 1(a) (saddle-shaped curve, with the four easy directions of magnetization having lower energy levels than the other principal crystallographic axes).



When two nanomagnets are placed next to each other, two types of dipole interaction arise. The first type arises when the magnetizations are perpendicular to the array axis. In this case, dipole interaction favors anti-parallel arrangement (or anti-ferromagnetic coupling) of adjacent magnetizations. The second type arises when the magnetizations are parallel to the array axis. In this case, dipole-dipole interaction favors a parallel alignment (or ferromagnetic coupling) of adjacent magnetizations. Using combinations of this dipole-dipole interaction and a small global static magnetic field to resolve tie situations, we realize a 2-bit NOR universal logic gate as explained below.

Consider the arrangement shown in Fig. 1 (b) through (e), where the two input nanomagnets, encoding bits AB and CD, are located on either side of the output nanomagnet encoding bits EF. Four different scenarios are investigated. The first scenario is shown in Fig. 1(b), where the magnetization direction of the input magnets is perpendicular to the axis of the nanomagnet array. When both input magnetizations are oriented up (or down) (first two rows of Fig. 1 b), the dipole coupling favors a down (or up) orientation of the output nanomagnet. When one input magnetization points up and the other points down (third and fourth rows of Fig. 1 b), the output nanomagnet is in a tied state, which we resolve using a global static magnetic field that forces the output magnetization to point "up". The second scenario is shown in Fig. 1(c), where input magnetizations are parallel to the nanomagnet array axis. When both input magnetizations are oriented left or right (first two rows of Fig. 1 c), the dipole coupling respectively favors left or right orientation of the output nanomagnet. When one input magnetization points left and the other points right (third and fourth rows of Fig. 1 c) the output nanomagnet is in a tied state, and ends up pointing neither left nor right, but orienting upward because of the global bias field. The third (Fig 1 d) and fourth (Fig 1 e) scenarios are mixed inputs, where one input magnetization



points perpendicular ("up/down") and the other parallel ("left/right") to the axis of the array, respectively favoring a "down/up" or "left/right" orientation of the output magnetization. These result in tied states. If one of the inputs is "up", the static bias field ensures that the output is "up", while if one of the inputs is "down", the bias field counters this, causing the output to point either "left" or "right" depending on the second input. The input (AB, CD) and output (EF) configurations, based on their binary bit representations, are transferred to a Karnaugh map (K-map) to simplify the logical expression of the output. The output table of the K-map is shown in Fig. 2. On simplification of the K-map, the result obtained (in Boolean notation) is $E = \overline{B}\overline{D}$ and $F = \overline{A}\overline{C}$, which is NOR logic.

We present detailed simulation results to show that the output magnetization always represents the NOR function of the inputs, independent of its initial orientation, when we clock the piezoelectric layer of the output multiferroic nanomagnet with a voltage to generate the proper stress cycle. To show this, we consider the total energy of the output nanomagnet when its magnetization vector subtends an angle $\theta_2$ with the positive x-axis[11]:

$$E_{total}(\theta_2) = \frac{\mu_0}{4\pi R^3}\left[M_s^2\Omega^2\right]\left[-2\cos\theta_2\left(\cos\theta_1 + \cos\theta_3\right) + \sin\theta_2\left(\sin\theta_1 + \sin\theta_3\right)\right] \quad (1)$$
$$+ \frac{K_1\Omega}{4}\sin^2 2(\theta_2 - \pi/4) - \frac{3}{2}\lambda_{100}\sigma\Omega\cos^2(\theta_2 - \pi/4) - \frac{\mu_0}{4\pi}\left[M_s\Omega\right]H_{applied}\sin\theta_2$$

where the first term is the dipole interaction energy with the neighbors subtending angles $\theta_1$ and $\theta_3$ with the positive x-axis, the second term is the magnetocrystalline anisotropy energy with $K_1$ being the first-order magnetocrystalline anisotropy constant, the third term is the stress anisotropy energy due to stress $\sigma$ applied along the [100] direction ($45^0$ with the x-axis) with $\lambda_{100}$ being the magnetostrictive constant in the direction of stress, and the last term is the interaction energy due to the static bias field $H_{applied}$ pointing in the "up" direction, or the [110] direction. Here, $\mu_0$ is the permeability of free space, $M_s$ is the saturation magnetization, $\Omega$ is the nanomagnet volume, and $R$ is the distance between the centers of two adjacent nanomagnets. The stress $\sigma$ is taken as positive for tension and negative for compression.

The energetic of such a system is briefly explained here. For single crystal nickel with $K_1 < 0$, the magnetic easy axes are <111> and hard axes are <100>. We tacitly assume that the two-dimensional geometry of the nanomagnet precludes out-of-plane magnetization orientation due to a large magnetostatic energy penalty. Thus, the magnetization is confined to the (001) plane and [110], [1$\bar{1}$0], [$\bar{1}\bar{1}$0] and [$\bar{1}$10] are the ground states which respectively correspond to the 90°, 0°, -90° and 180° orientations in Fig. 1. The dipole and static magnetic field interaction energies are not high enough to move the magnetization away from these minima. However, upon applying a stress, the magnetizations are pushed out of these minima. Since the magnetostrictive coefficient $\lambda_{100}$ of Ni is negative, a tensile stress along [100] rotates the magnetization to either the -45° or the +135° state (depending on which is closest to its initial state), while a compressive stress along [100] direction rotates the magnetization to either the -135° or +45° states. When this stress is released, the dipole interactions and the static bias field determine which of the two adjacent easy directions the output magnetization settles into. *Thus, to rotate the magnetization through 180° one needs both a tensile and compressive stress cycle; with each half-cycle producing a +90° rotation.*

For numerical simulation, the multiferroic nanomagnets were assumed to be made of two layers: single crystal nickel and lead-zirconate-titanate (PZT) with the following properties for Ni: $\lambda_{100} = -2 \times 10^{-5}$, $K_1 = -4.5 \times 10^3$ J/m$^3$, $M_s = 4.84 \times 10^5$ A/m [12], and Young's modulus $Y = 2 \times$ $10^{11}$ Pa. The PZT layer can transfer up to $500 \times 10^{-6}$ strain to the Ni layer[13]. The nanomagnets are assumed to be circular disks with a diameter $d$ = 100 nm and thickness = 10 nm, while the center-to-center separation (or pitch) is 160 nm. The above parameters were chosen to ensure that: (i) The magnetocrystalline energy barrier of the nanomagnets is sufficiently high (~0.55 eV or ~22$kT$ at room temperature) so that the static bit error probability due to spontaneous magnetization flipping is very low; (ii) The stress anisotropy energy (~1.5 eV) generated in the magnetostrictive Ni due to a strain of $500 \times 10^{-6}$ transferred from the PZT layer can clearly rotate the magnetization against the magnetocrystalline anisotropy; (iii) The dipole interaction energy is limited to 0.2 eV, which is lower than the magnetocrystalline anisotropy energy. This prevents the magnetization from switching spontaneously without the application of the electric-field induced stress for clocking.

To show that the "output" nanomagnet behaves as desired (for the various configurations shown in Fig. 1), its total energy ($E_{total}$) is computed as a function of $\theta_2$ upon application of the stress clock cycle tension → relaxation → compression → relaxation, in increments/decrements of 0.1 MPa up to a maximum amplitude of 100 MPa. For each value of the stress, the local energy minimum closest to the previous state determines the final magnetization orientation $\theta_2$. We study a particular case: input nanomagnets having magnetization directions as 'right' and 'up', i.e. AB = '01' and CD = '00' with the output magnet 'EF' initially in the 'down' ('11') state as shown in Fig 3. All other cases are exhaustively studied in the supplementary material that accompanies this letter (see Figs S1 to S10 in the supplement, Ref 14). When tensile stress is applied to the nanomagnet along the [100] direction (i.e. at $\theta = 45°$) the output magnetization rotates right and settles at -45° as shown in Fig 3 (a). This is because Ni has negative magnetostriction and a tensile stress tends to rotate the magnetization away from the 45⁰ stress axis. Of the two perpendicular directions (+135° and -45°), -45° is closer and hence the magnetization rotates from -90° to -45°. In the next stage, the stress on 'EF' is stepped down to zero. The result, shown in Fig. 3(b), indicates relaxation of the magnet's magnetization from -45° to 0° as expected. This can be explained by understanding the effect of the bias field and dipole interaction on the output magnet. The left input (AB) favors the output magnetization orienting parallel to it, i.e. pointing right, while the right input (CD) favors the output aligning anti-parallel to it, i.e. pointing down. However, the global bias field, pointing up, resolves the tie between the "down" (or -90º) state and "right" (or 0 º) state in favor of the "right" (or 0 º) state, causing the output magnetization vector to settle to 0° when stress is relaxed to zero. Following this, a compressive stress is applied to 'EF' at +45° that rotates the magnetization from 0° to +45° (Fig. 3(c)). Subsequently, when stress on 'EF' is relaxed to zero, the final state of the output magnet settles back to 0° ('right') under the influence of dipole interaction and the bias field as expected. Thus at the end of the cycle, the output EF = '01' is realized, showing successful NOR operation.

For universality, the initial state of the output nanomagnet should not affect the desired result. To verify this, we performed simulations for the input combination AB = '01', CD = '00' with the other three possible initial orientations of the output magnet (see Fig. S1, S2 and S3 in the supplement, Ref 14) and have shown that the final output state EF = '01' is achieved regardless of its initial orientation. Simulation of the output for all other input gate combinations are exhaustively covered in the supplementary material (see Fig. S4 - S10, Ref 14).

In conclusion, we have shown the feasibility of four state nanomagnetic logic using multiferroic nanomagnets. It is obvious that if the initial state of any nanomagnet is not in one of the four stable states, it will relax to the nearest stable state, thus behaving like an associative memory element[15]. They have applications in pattern recognition and other signal processing.

**Figure Captions**

**Fig. 1 (a):** Ni/PZT multiferroic nanomagnet with single crystal Ni magnetostrictive layer having a biaxial anisotropy that creates four possible magnetization directions (easy axes) – 'up' (00), 'right' (01), 'down' (11) and 'left' (10). The bit assignments ($AB$, $A\bar{B}$, $\bar{A}B$, $\bar{A}\bar{B}$) are also shown. The saddle-shaped curve represents the energy profile of the nanomagnet in the ground/unstressed state, in which the energy minima are located along the easy axes. As a result, each of these four axes is a possible direction for the magnetization. Nanomagnet array with dc bias magnetic field showing NOR logic realization for all input combinations. The two input nanomagnets (AB, CD) are placed on either side of the output magnet (EF). The dotted arrows indicate the occurrence of a tie-condition (output has two equally possible choices) that is resolved by the influence of the bias field's direction. (b) The input combinations have magnetization directions perpendicular to the magnet array axis, resulting in the output direction having two possible orientations – 'up' or 'down' (c) The magnetization direction of the inputs are parallel to the magnet axis. Consequently, the output orientations are either 'left' or 'right' or 'up (tie condition)'. (d) The left input magnet, AB, is either 'left' or 'right' while the right input, CD, is either 'up' or 'down'. The output is, therefore, either 'left' or 'right' for non-tie cases and 'up' (determined by bias field) when a tie condition arises. (e) AB is either 'up' or 'down' while CD is 'left' or 'right'. Similar to (d), the outputs are either 'left', 'right' or 'up (tie condition)'.

**Figure 2:** A Karnaugh-map representation of the input (AB, CD) combinations is illustrated, with the output EF indicated in the dotted rectangle, which is then separated into individual 'E' and 'F' sub-K-maps in order to determine their logical expressions $E = \overline{BD}$ and $F = \overline{AC}$.

**Figure 3:** Energy plots of the output nanomagnet (EF) as a function of the magnetization angle. The initial conditions used are: AB = '01', CD = '00' and EF = '11'. (a) With no stress applied to magnet EF initially, the magnetization direction begins at -90°. The first stage of the stress cycle (Tension at +45°) is then initiated, which causes the magnetization to rotate away from the stress axis to the closest energy minima (-45°). (b) Upon relaxation of stress on EF (stage 2), the closest energy minimum is at 0° and therefore, the magnetization rotates into that position. (c) Stage 3 of the stress cycle involves the application of a compressive stress (negative, +45°) on EF. The energy minima are located along the stress axis; therefore, the magnetization rotates and settles at +45°. (d) Finally, the stress on EF is relaxed which causes the magnetization to rotate to the closest energy minimum (0°). The arrows indicate the direction of applied stress and the resulting magnetization rotation.

**List of References**

1. R. P. Cowburn and M E Welland, Science, **287**, 1466-1468 (2000).

2. B. Behin-Aein, S. Salahuddin and S. Datta, IEEE Trans. Nanotech., IEEE Trans. Nanotech., **8**, 505 (2009).

3. G. Csaba, A. Imre, G. H. Bernstein, W. Porod and V. Metlushko, IEEE Trans. Nanotech., **1**, 209 (2002).

4. B. Behin-Aein, D. Datta, S. Salahuddin and S. Datta, Nature Nanotech., **5**, 266 (2010).

5. M. T. Alam, M. J. Siddiq, G. H. Bernstein, M. Niemier, W. Porod and X.S Hu , IEEE Trans. Nanotech., **9**, 348 (2010).

6. D.C. Ralph, M.D. Stiles, J. Magn. Mag. Mat., **320,** 1190 (2008).

7. J. Atulasimha and S. Bandyopadhyay, Appl. Phys. Lett.**, 97,** 173105, (2010).

8. M. S. Fashami, K. Roy, J. Atulasimha and S. Bandyopadhyay, arXiv:1011.2914v1.

9. K. Roy, S.Bandyopadhyay and J. Atulasimha, arXiv:1012:0819v1.

10. S. A. Wolf, J. Lu, M. R. Stan, E. Chen and D. M. Treger, Proc. IEEE, **98**, 2155 (2010).

11. S. Chikazumi, *Physics of Magnetism*, Wiley New York, 1964.

12. E. W. Lee, Rep. Prog. Phys., **18**, 184 (1955).

13. M. Lisca, L. Pintilie, M. Alexe and C.M. Teodorescu, Appl. Surf. Sci., **252**, 30 (2006).

14. Supplementary material located at…..

15. V.P. Roychowdhury, D. B. Janes, S. Bandyopadhyay and X. Wang, IEEE Trans. Elec. Dev., **43**, 1688 , (1996); N. Ganguly, P. Maji, B. Sikdar and P. Chaudhuri, IEEE Trans. Syst. Man Cyber., Part B, **34**, 672, (2004).

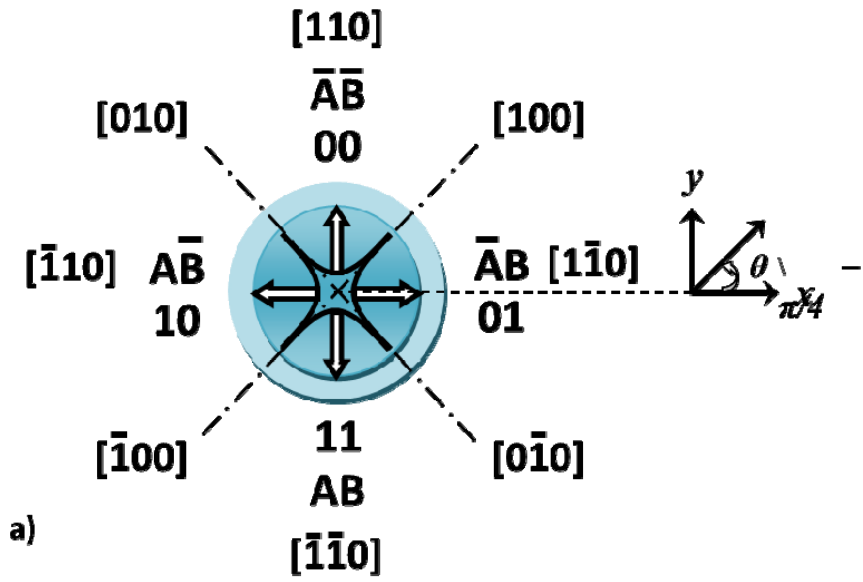

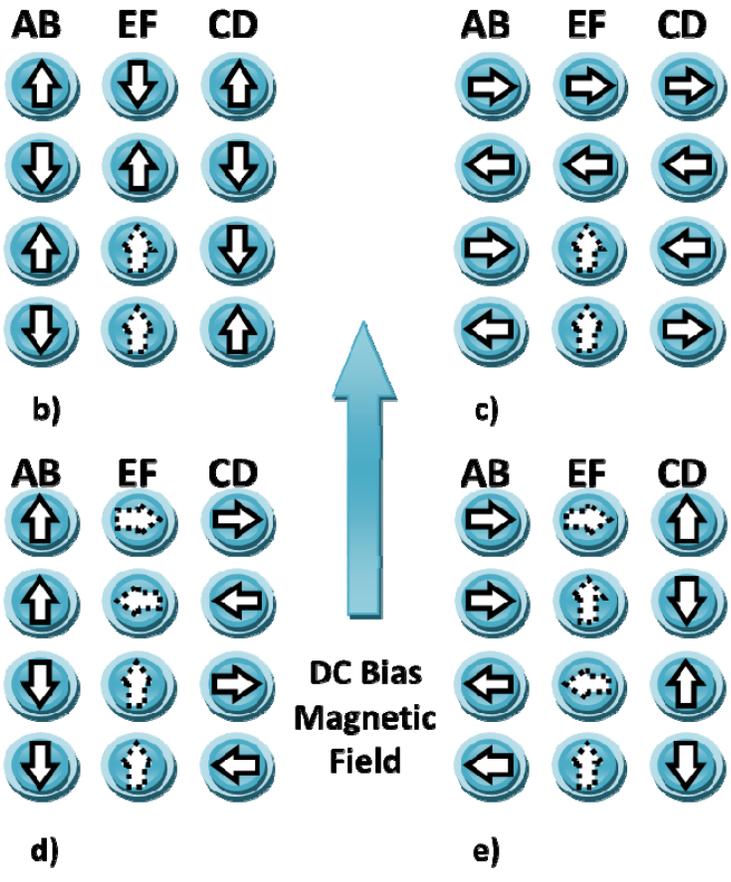

**Fig. 1**

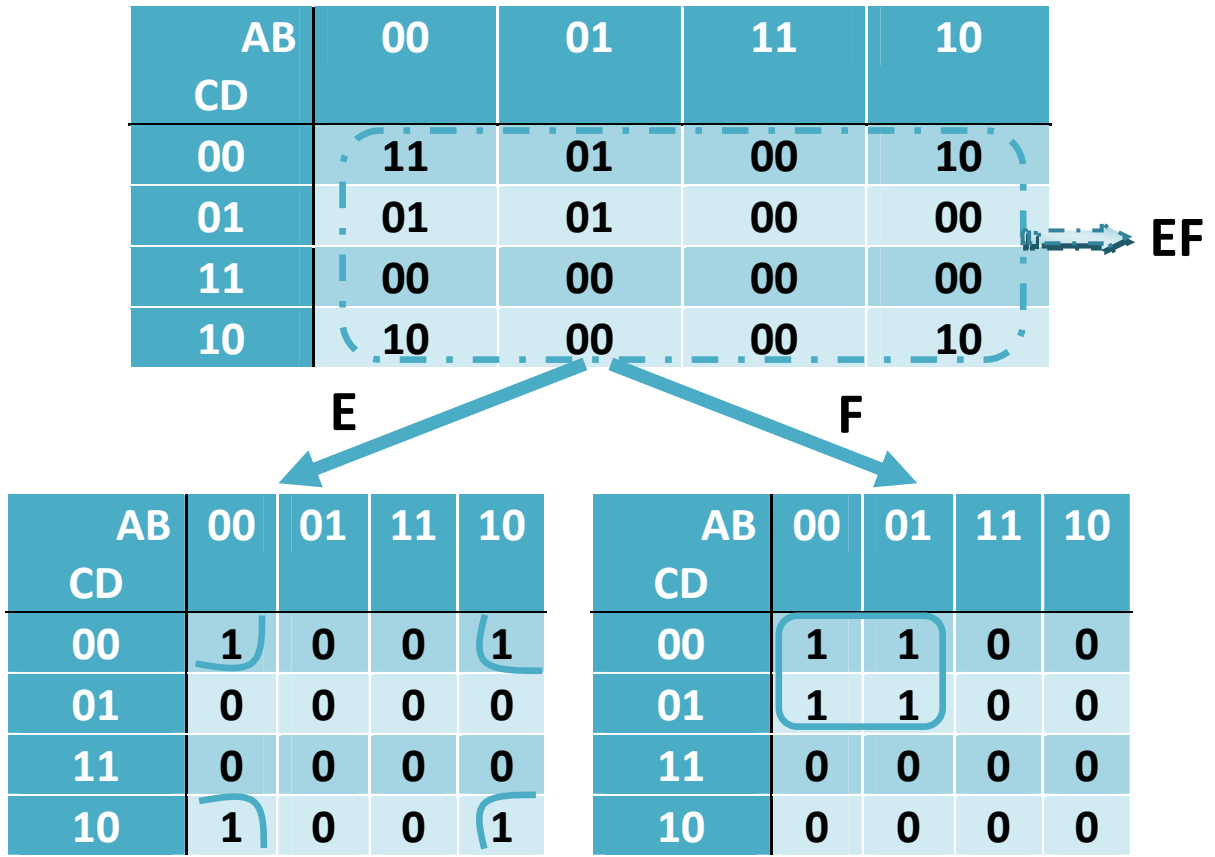

Fig. 2

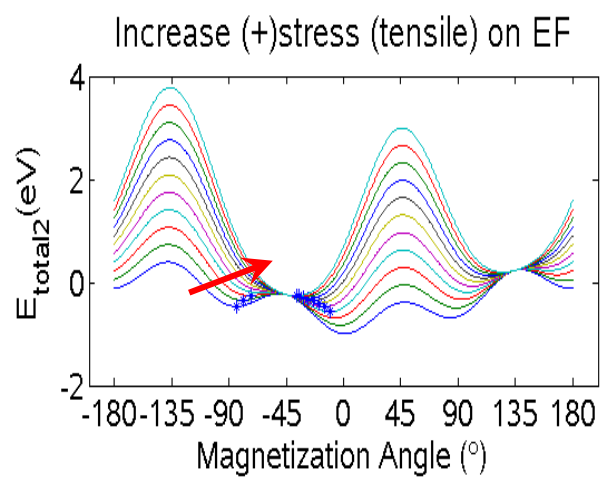

**a)**

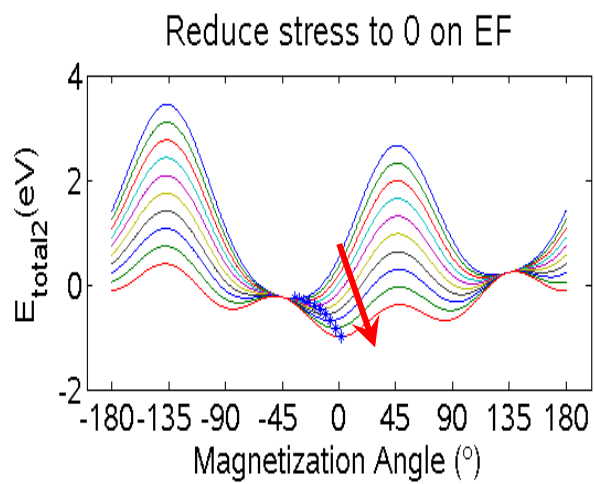

**b)**

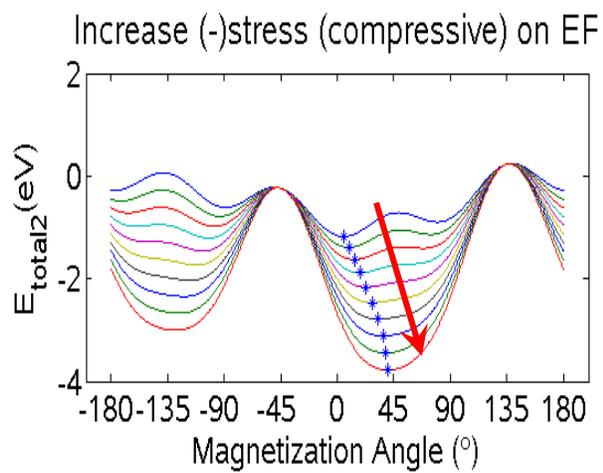

**c)**

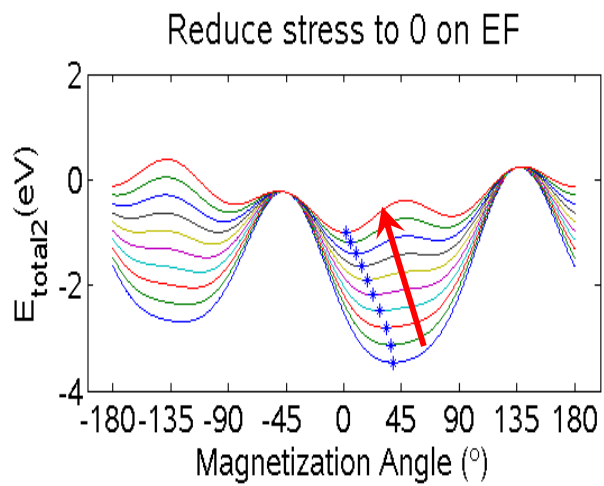

**d)**

**Fig. 3**

# SUPPLEMENTARY INFORMATION

# FOUR-STATE NANOMAGNETIC LOGIC USING MULTIFERROICS

Noel D'Souza[a], Jayasimha Atulasimha[a], Supriyo Bandyopadhyay[a]

Email: {dsouzanm, jatulasimha, sbandy}@vcu.edu

(a) Department of Mechanical and Nuclear Engineering,

(b) Department of Electrical and Computer Engineering,

Virginia Commonwealth University, Richmond, VA 23284, USA.

In the letter, we showed that dipole-coupled Ni/PZT multiferroic nanomagnets with binary bits encoded in the four stable magnetization directions ('up', 'down', 'left' and 'right') can implement 4-state NOR logic with the proper clock sequence. We showed this for one particular input bit combination and for one particular initial state of the output. For the sake of completeness, we have to show it for all possible input combinations and for all possible initial state of the output magnet. This is shown here.

Of the sixteen possible input configurations (input bits) and four possible initial states of the output magnet (output bit), the results and energy plots for one particular input configuration (AB = 'right', CD = 'up') and one initial state of the output (magnetization direction pointing 'down') was shown in the letter. In this supplement, we consider the other cases in order to be exhaustive. We first pick the input configuration (AB = 'right', CD = 'up') discussed in the letter and show that for this input combination, the final state of the output is independent of the initial state of the output. There are three other possible initial states of the output ('left', 'right' and 'up') and each of them is examined for the above input combination (Figs. S1 – S3). In each case, the final output settles in the correct direction ('right') conforming to NOR logic. Thus, the final output is independent of the initial state of the output for this input combination. This means that the output is determined solely by the inputs and hence is a unique, single-valued function of the inputs.

We also show the results obtained from the seven other unique input combinations when the initial state of the output is EF = 11 (Figs. S4 – S10). It is obvious that the final state of the output will be independent of the initial state for these input combinations as well. The remaining eight input combinations are not examined since they are equivalent to ones examined here due to symmetry. For all input combinations, the NOR function is always realized.

**Supplement Figure Captions**

**Figure S1:** Energy plots of the output magnet representing the bits EF with input bits AB = '01', CD = '00' and EF = '00' initially. (a) Rotation from +90° to +135° as a consequence of tension applied along the +45° axis. (b) Upon relaxation of the stress, the magnetization vector rotates back to +90°. (c) With compression applied on the output magnet along the +45° axis, its magnetization rotates to +45°. (d) Finally, when the stress is relaxed to zero, the output magnet rotates its magnetization to 0°, completing the NOR logic operation.

**Figure S2:** The input combinations AB='01', CD='00' are the same as in figures S1. The initial state of the output magnet EF is set at '01'. The stress cycle (tension, relaxation, compression, relaxation) is applied to the output magnet with the magnetization rotating sequentially through -45°, 0°, +45° and finally settling at 0°, thereby once again completing the NOR operation.

**Figure S3:** AB = '01', CD = '00' as in figure S1. The initial output state is set to EF = '10'. The stress cycle applied to the output magnet causes its magnetization to rotate sequentially through +135°, +90°, +45° and 0°, thereby completing the NOR operation. Figs. S1-S3 show that the final state of the output is indeed independent of the initial state and hence determined uniquely by the inputs.

**Figure S4:** AB = '00', CD = '00' and EF = '11'. Since the inputs are pointing 'up', the dipole interaction pushes the output magnet's magnetization vector 'down'. The stress cycle applied to output magnet causes its magnetization vector to rotate sequentially through -45°, -90°, -135° and back to -90°, completing the NOR operation.

**Figure S5:** AB = '00', CD = '11' and EF = '11'. The input magnetization directions, on either side of the output magnet encoding EF, point in opposite directions ('up' and 'down'). As a result, the dipole interaction of the inputs on the output cancels out. This would result in a tie-condition when the stress cycle is applied (specifically, during the relaxation phases, when the magnetization would have two equally likely directions to settle into). However, when a dc bias magnetic field is applied [$H_{appl}$ = 1000 A/m (~12 Oe)], the energy profile is no longer symmetric and is slightly biased towards +90°. Now, when the stress cycle is applied to the output magnet, its magnetization rotates through +135°, +90°, +45° and 90°, thereby once again completing the NOR operation.

**Figure S6:** AB = '01', CD = '01' and EF = '11'. In this case, the inputs are both pointing towards the 'right'. Hence, the dipole interaction shows a strong preference for ferromagnetic

coupling (parallel arrangement). This can be seen in the energy profile of the output magnet, EF, which has an absolute energy minimum located at 0° ('right'). The magnetization rotation arising due to the stress cycle applied to the output magnet is from the initial -90° direction to -45° (*since the ferromagnetic coupling due to the dipole interaction is strong, the magnetization easily rotates to 0° at low values of applied stress. However, at higher stress values, the stress anisotropy energy is greater than the dipole energy and, consequently, the magnetization settles at -45°, 0°, +45° before settling to 0°*. Ultimately, the NOR operation is once again realized.

**Figure S7:** AB = '01', CD = '10' and EF = '11'. Since the input magnetizations point in opposite directions, there is no net dipole interaction on the output magnet encoding the bits EF (similar to the configuration of Fig. S5). Once again, the applied bias magnetic field tips the energy profile of the output magnet towards +90°. The stress cycle applied to the output magnet causes its magnetization to rotate through -45°, 0°, +45° and +90°, thus implementing the NOR operation.

**Figure S8:** AB = '01', CD = '11' and EF = '11'. In this configuration, AB points 'right' while CD points 'up'. In the first stage of the stress cycle (tension along +45°) on the output magnet, the magnetization rotates from -90° to -45° (similar to Fig. S6, at low tensile stresses, the preferred alignment is 0°, parallel to AB. Further increases in stress cause the magnetization to settle at -45°). Relaxation of the stress then rotates it to 0°, compression takes it to +45° and ultimately, relaxation causes it to settle at +90°. The NOR operation is realized.

**Figure S9:** AB = '10', CD = '00' and EF = '11'. With AB pointing 'left' and CD pointing 'up', the dipole interaction is similar to that of the case in Fig. S7, with the output EF preferring a parallel alignment with AB. The stress cycle applied to the output magnet causes its magnetization to rotate from the initial direction of -90° to -45°, -90°, -135° and finally, -180°, thus implementing the NOR function.

**Figure S10:** AB = '10', CD = '11' and EF = '11'. This configuration (AB points 'left', CD points 'down') is similar to that of Fig. S8. The applied bias field tries to align the output magnet's magnetization vector along the +90° direction, without which a tie-condition would arise (two equally possible directions). The stress cycle applied to the output magnet causes its magnetization to rotate sequentially through -45°, 0°, +45° and +90°. This completes the NOR operation.

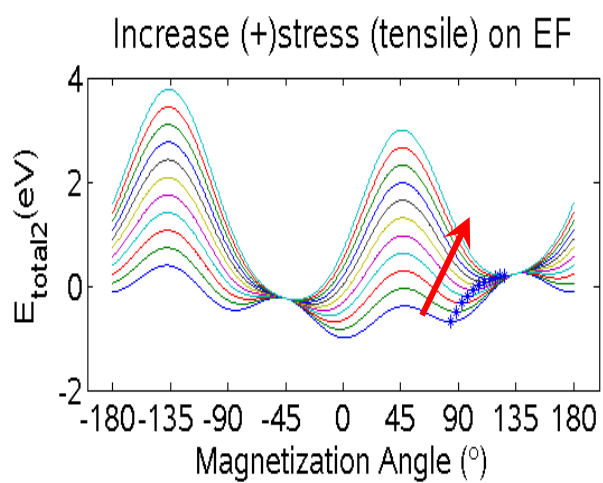

a)

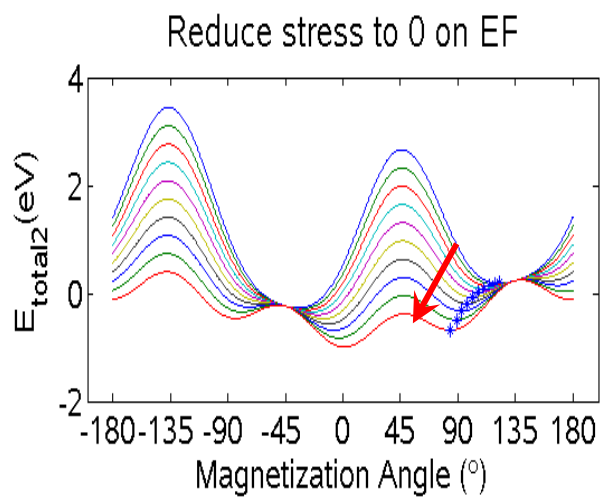

b)

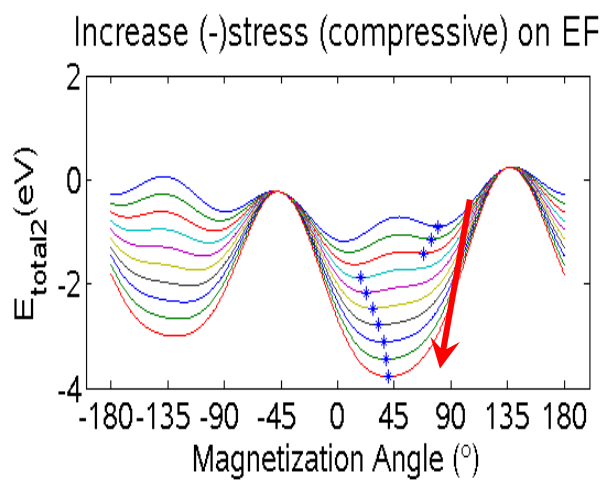

c)

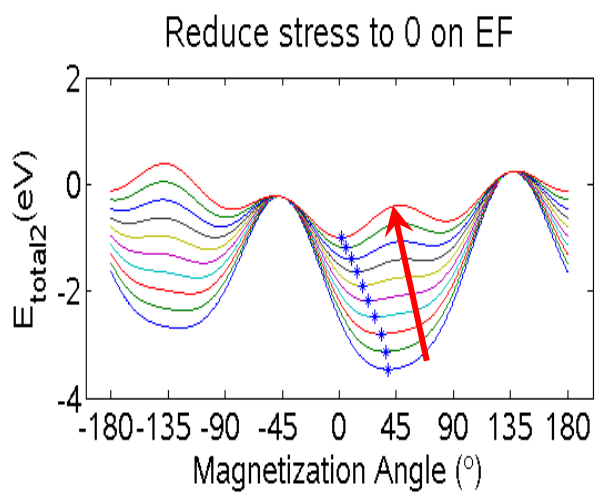

d)

**Fig. S1**

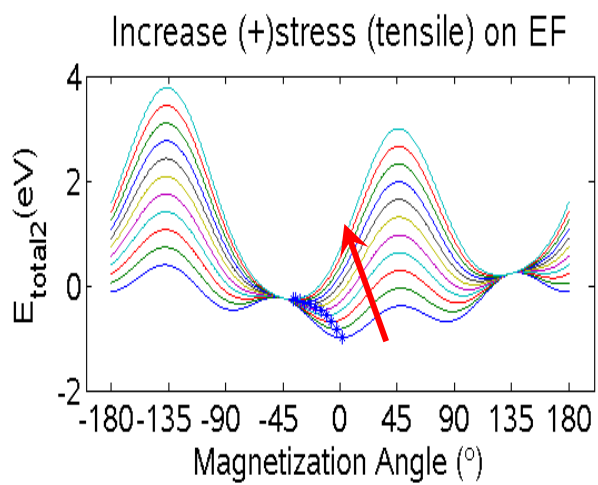

a)

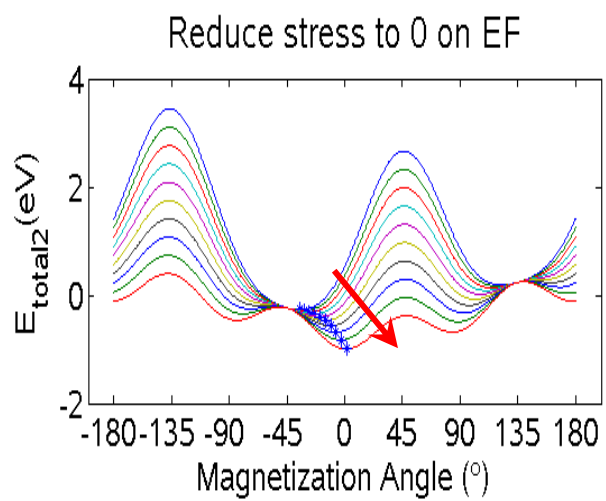

b)

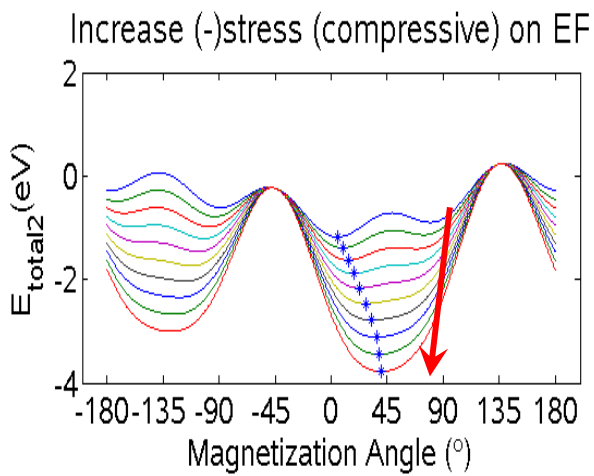

c)

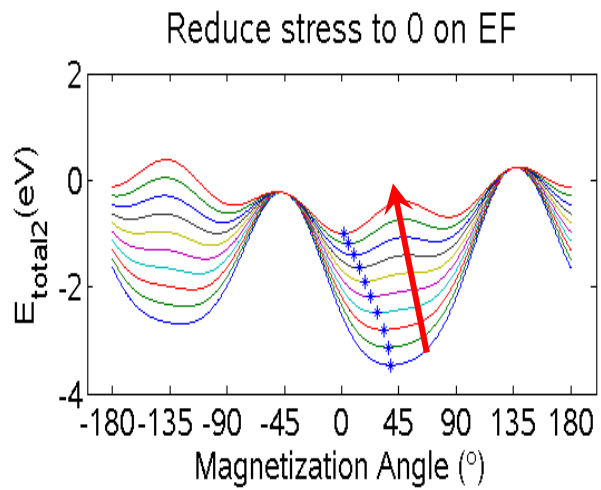

d)

**Fig. S2**

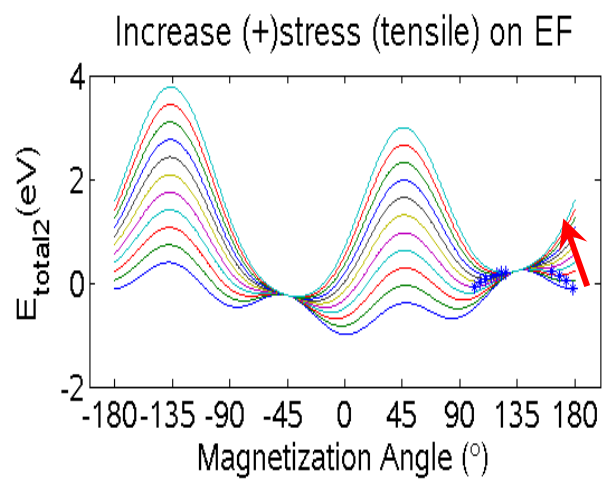

a)

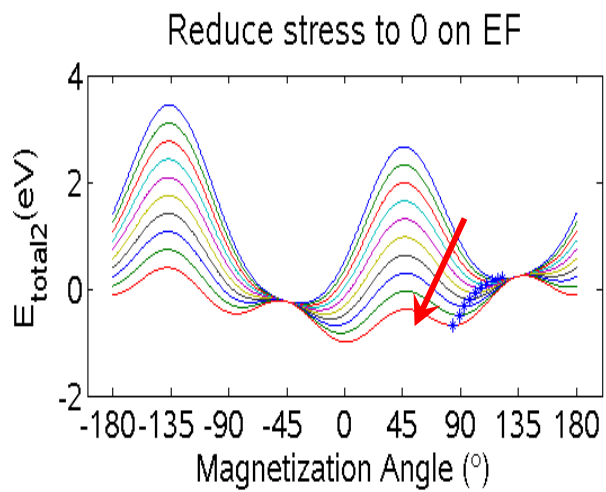

b)

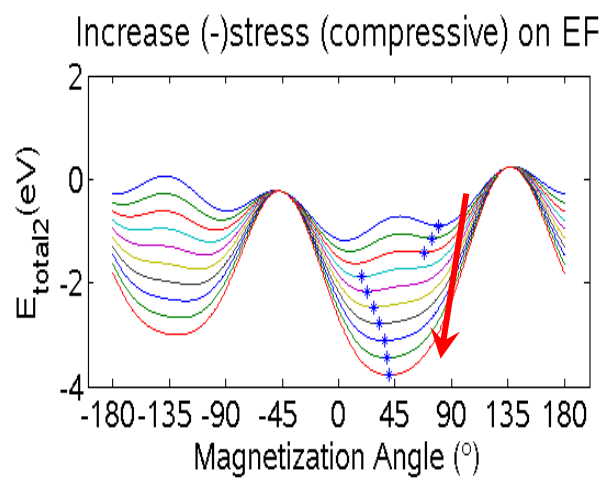

c)

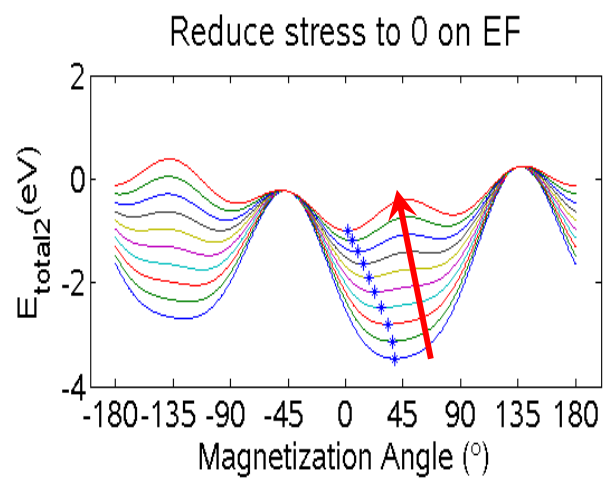

d)

**Fig. S3**

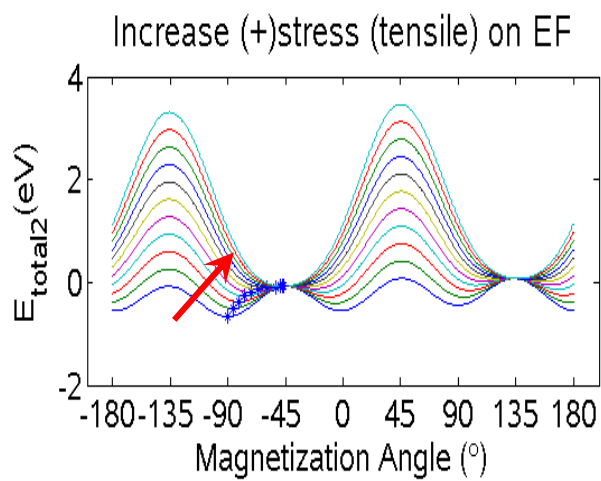

a)

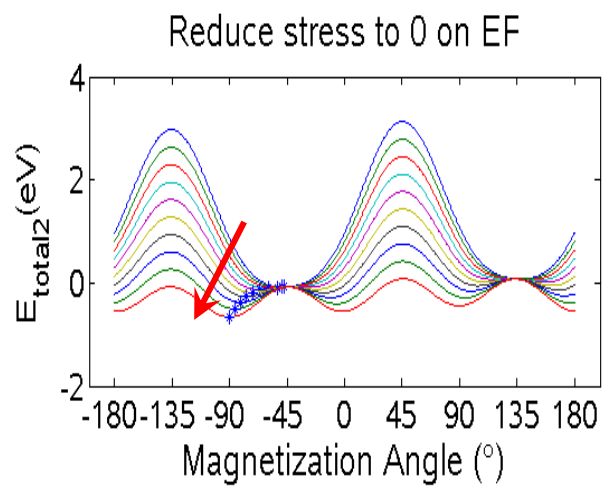

b)

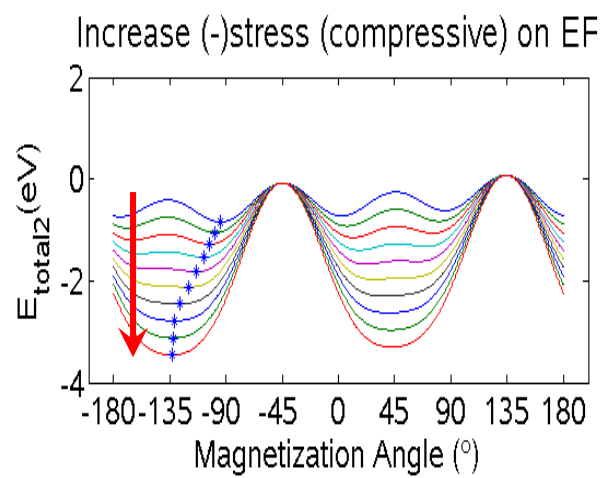

c)

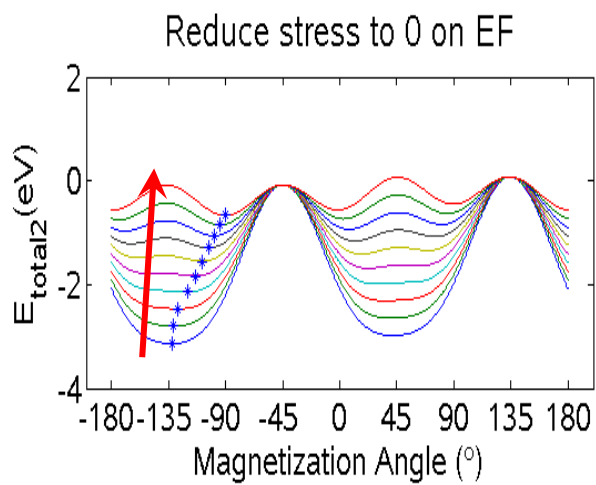

d)

**Fig. S4**

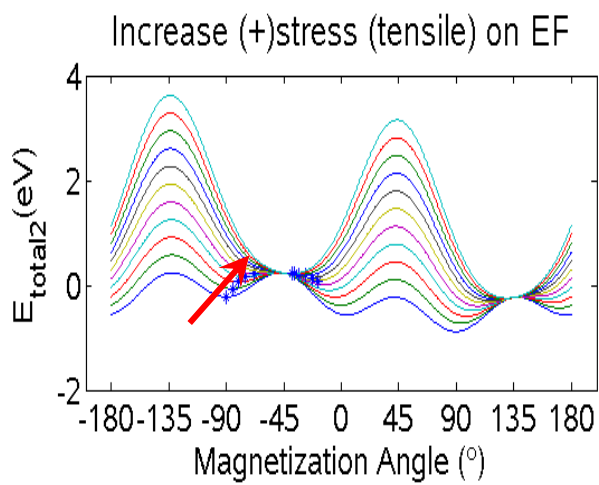

a)

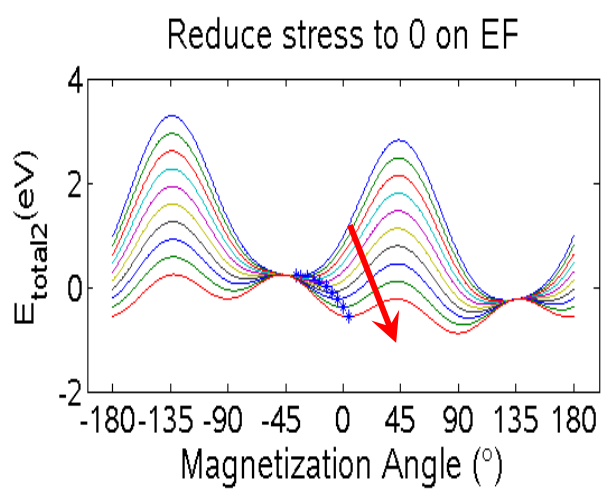

b)

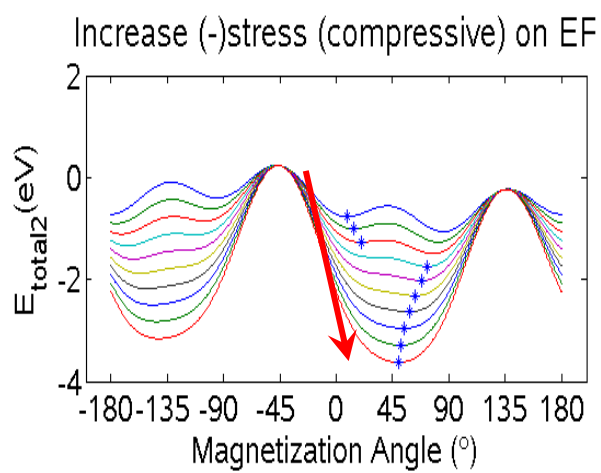

c)

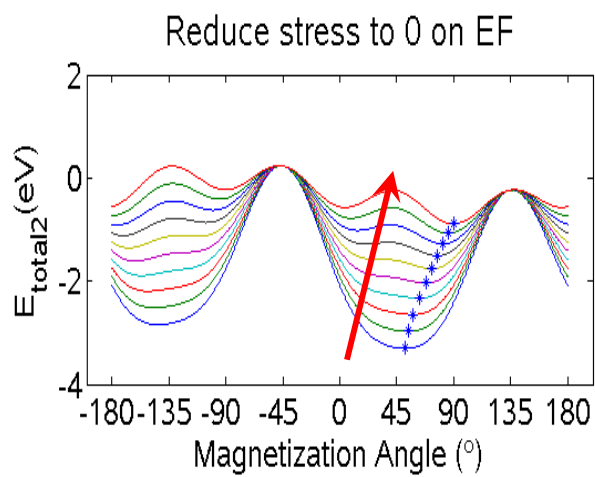

d)

**Fig. S5**

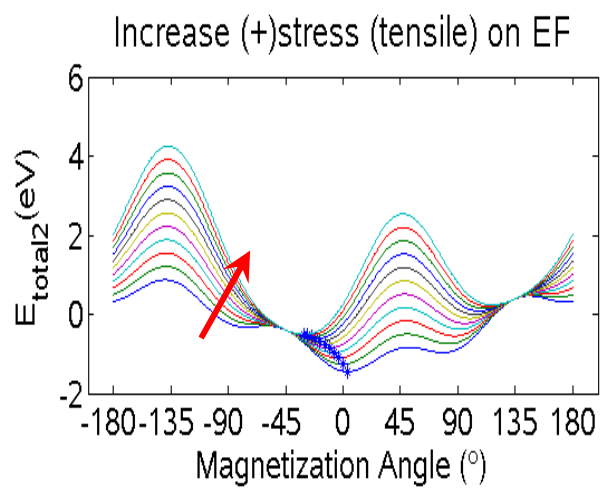

a)

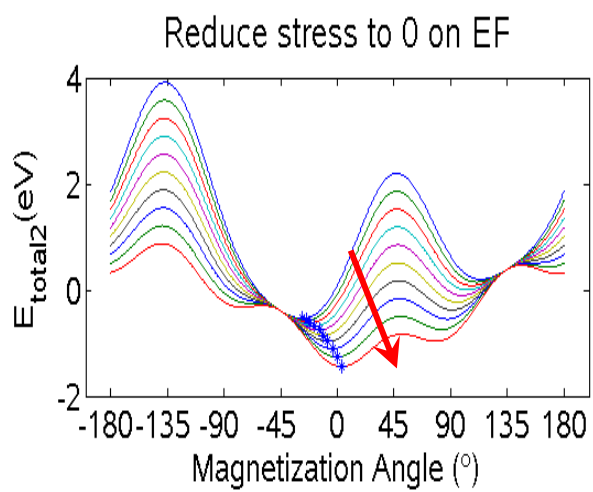

b)

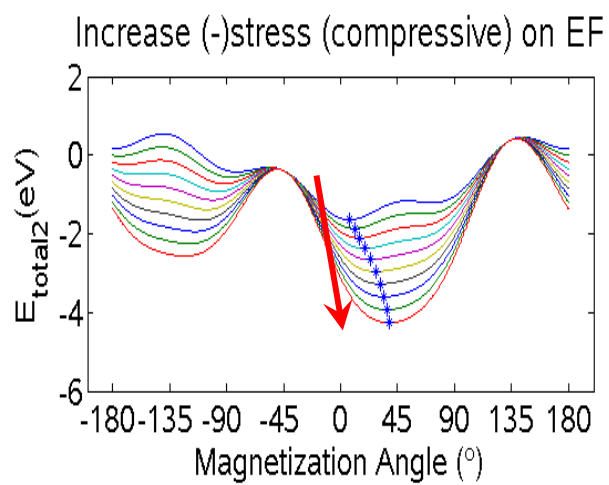

c)

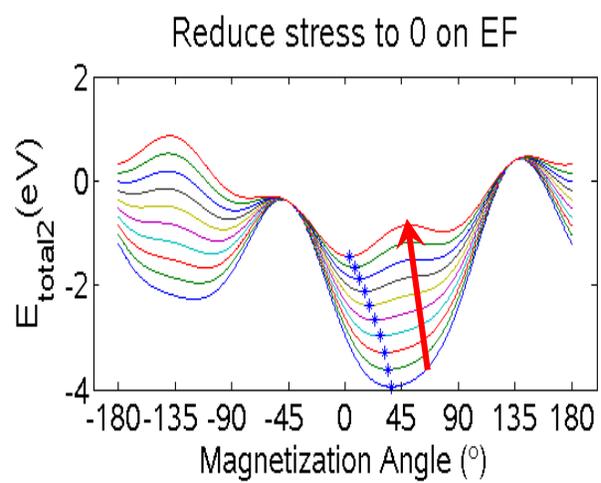

d)

**Fig. S6**

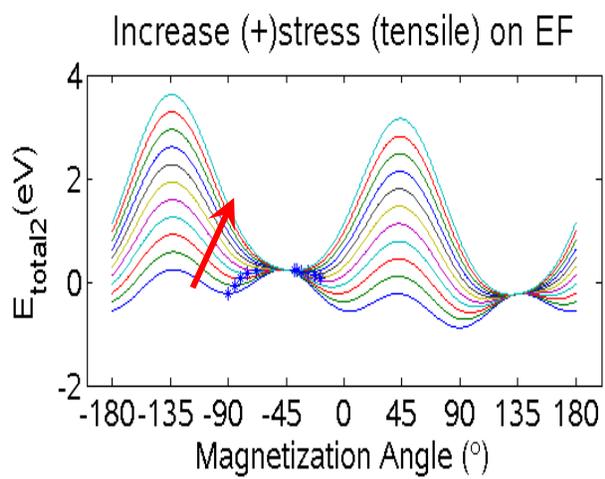

a)

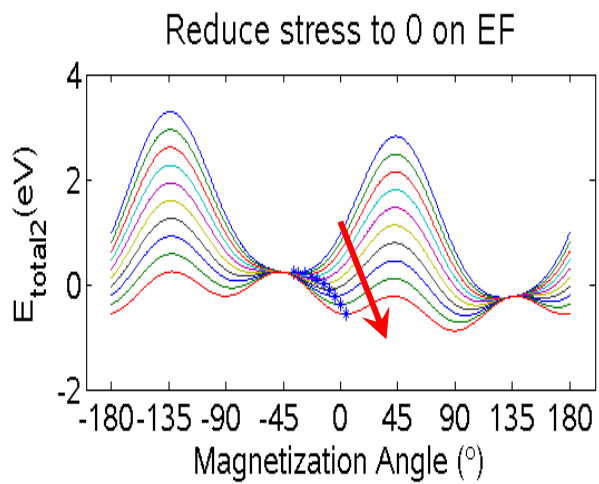

b)

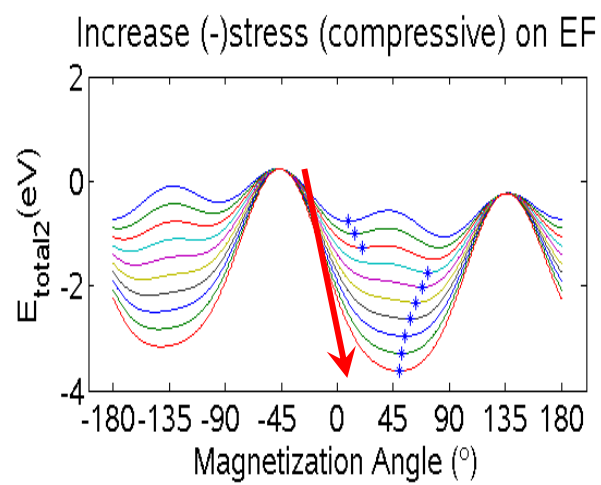

c)

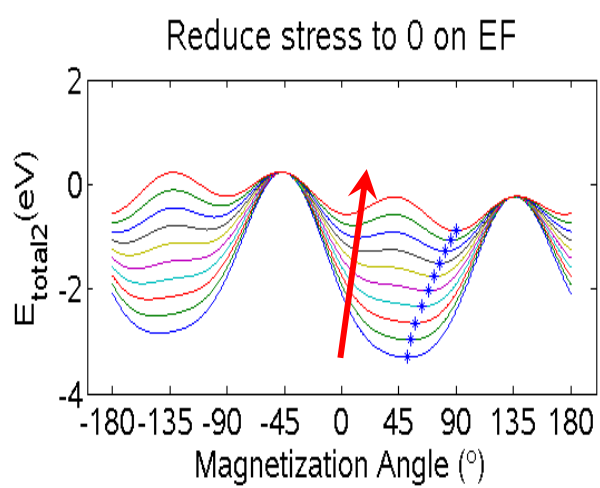

d)

**Fig. S7**

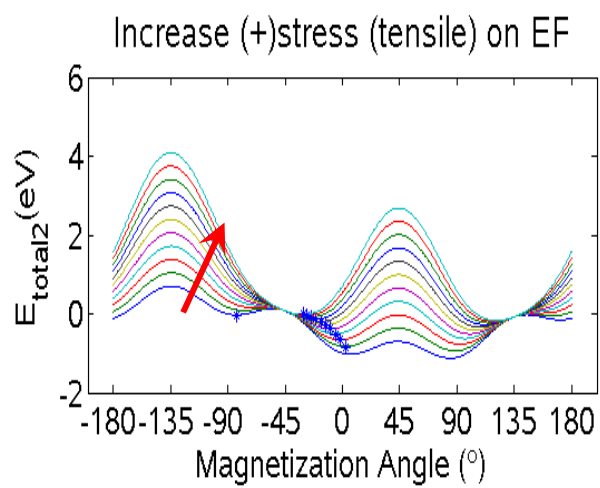
a)
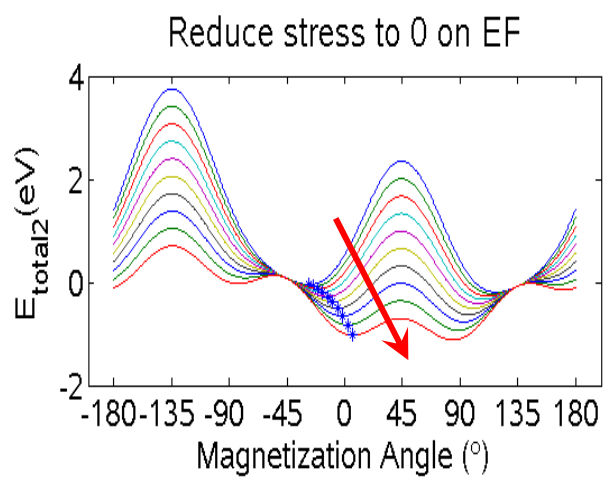
b)
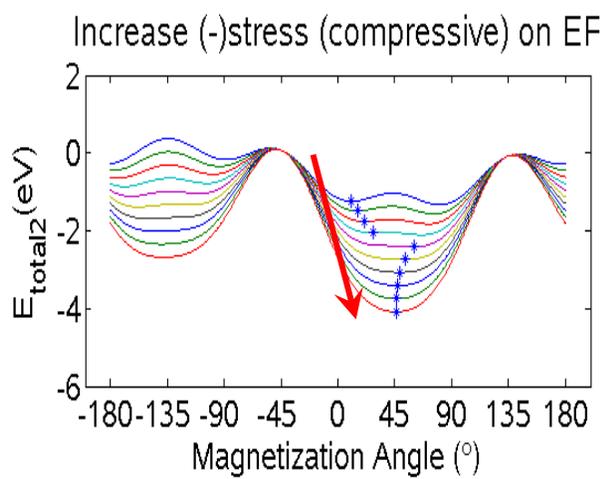
c)
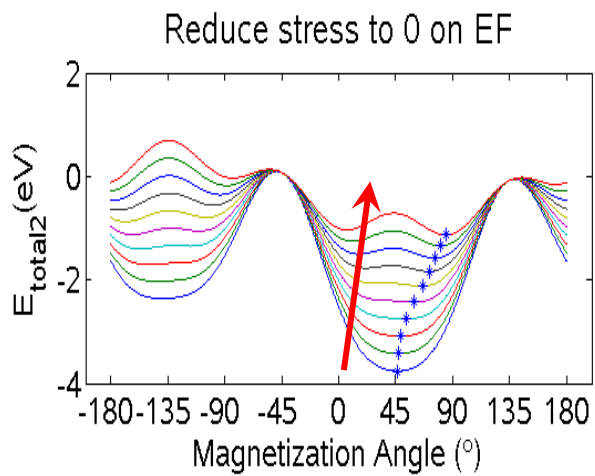
d)

**Fig. S8**

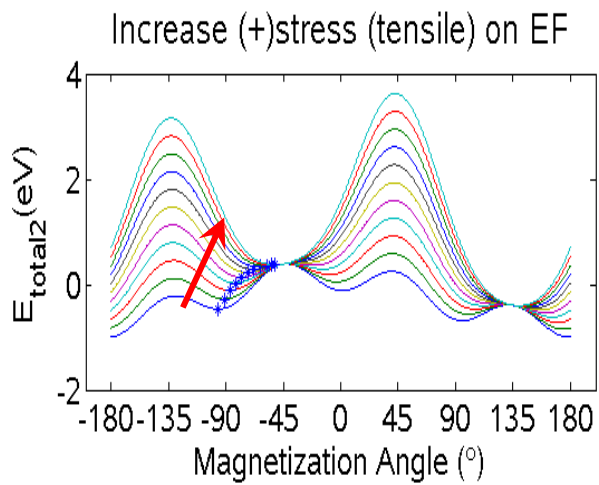

a)

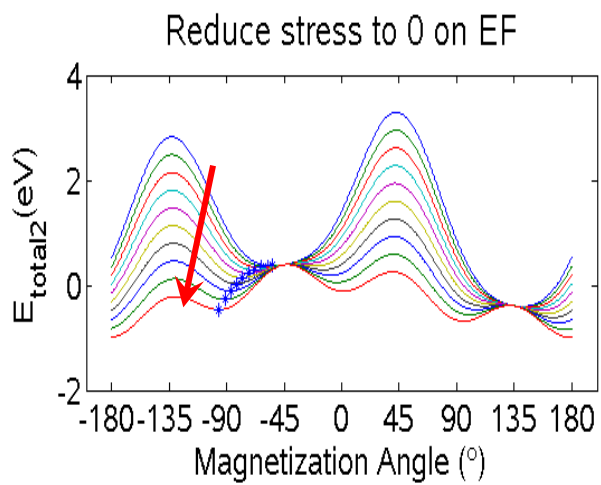

b)

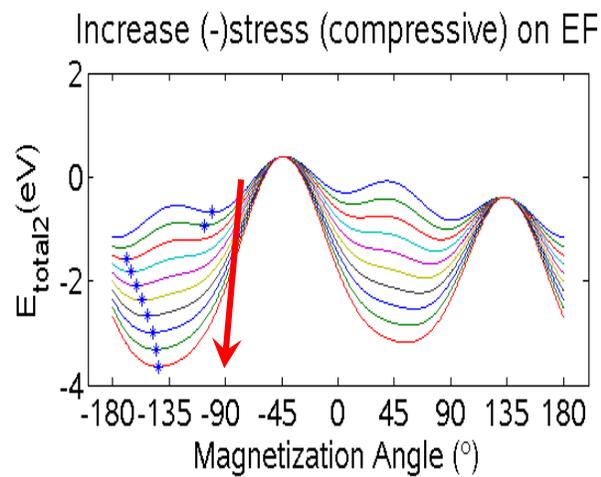

c)

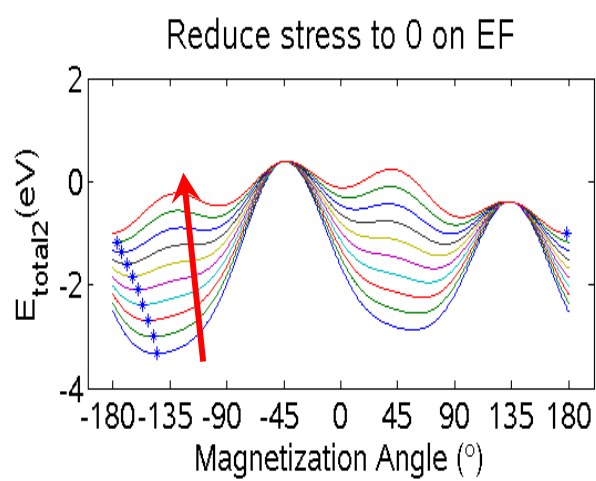

d)

**Fig. S9**

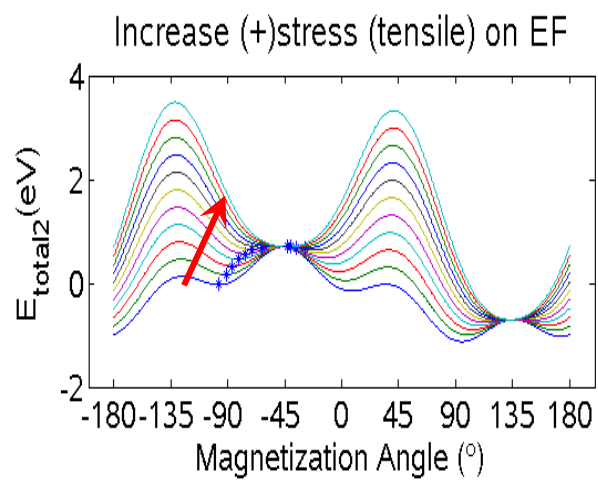

a)

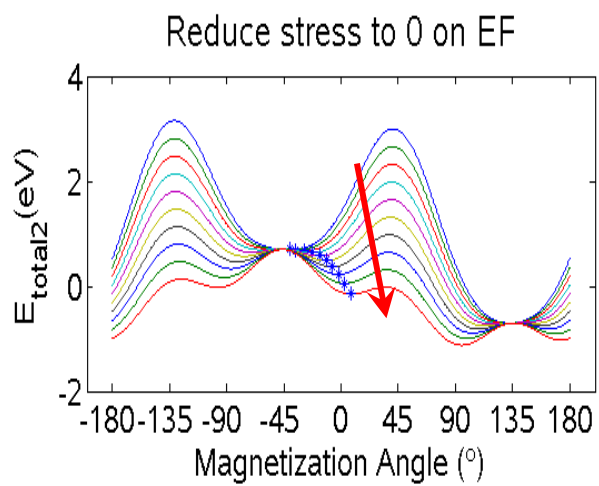

b)

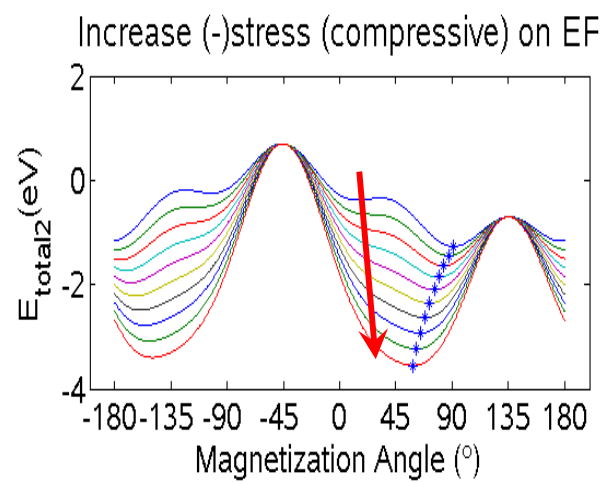

c)

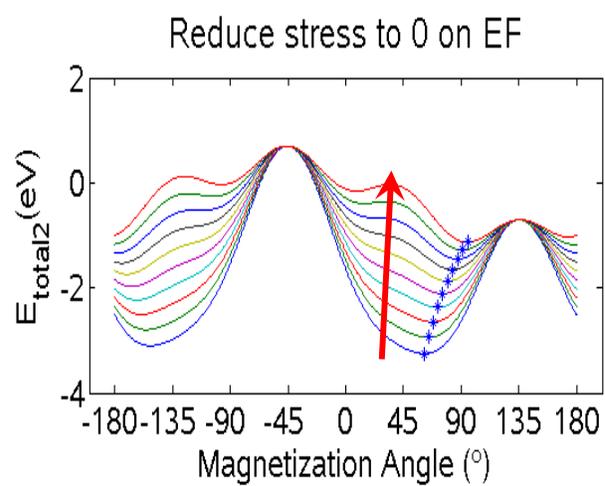

d)

**Fig. S10**